\documentclass{optica-article}

\journal{opticajournal} 

\articletype{Research Article}

\usepackage{lineno}
\usepackage{subcaption}
\usepackage{amsmath}
\usepackage{booktabs, siunitx}

\usepackage{xcolor}

\def\ls{\ell_\text{s}}
\def\lt{\ell_\text{t}}

\begin{document}

\title{Quasiannealed Monte Carlo method for light transport in strongly heterogeneous media}

\author{Lo\"{i}c Tran,\authormark{1,2} Benjamin Askenazi,\authormark{2} and Kevin Vynck\authormark{3,*}}

\address{\authormark{1}Laboratory for Photonics Numerics and Nanosciences (LP2N), University of Bordeaux, IOGS, CNRS, 1 rue François Mitterrand, 33400 Talence, France\\
\authormark{2}L'Or\'{e}al Research and Innovation, 9 rue Pierre Dreyfus, 93400 Clichy, France\\
\authormark{3}Institute of Light and Matter (iLM), Claude Bernard University Lyon 1, CNRS, 10 rue Ada Byron, 69622 Villeurbanne Cedex, France}

\email{\authormark{*}kevin.vynck@univ-lyon1.fr}

\begin{abstract*} 
Random-walk Monte Carlo simulations are widely used to predict the optical properties of complex, disordered materials. In presence of large heterogeneities (e.g., spatially-extended nonscattering regions in a turbid environment), an explicit description of the micro and macrostructures and of the light propagation therein is generally required, in addition to a statistical average over a representative set of microstructures, thereby making simulations in so-called ``quenched'' disorder particularly time-consuming.
We explore here the possibility to model light transport in finite-size strongly heterogeneous media without an explicit description of the underlying microstructure but from the knowledge of typical random-walk trajectories in infinite-size media, that take correlations between successive interaction events into account. Simulations may thus be performed for media of any macroscopic shape and size more efficiently. We illustrate this approach, coined ``quasiannealed'', with the case of a two-phase emulsion consisting of transparent spherical droplets dispersed in a turbid medium. Good agreement with predictions from simulations in quenched disorder on the reflectance of finite-thickness slab is found for a large set of microstructure properties and thicknesses with typical errors on the reflectance on the order of a percent.
\end{abstract*}

\section{Introduction}
\label{sec:introduction}

Light transport in disordered media plays a central role in many fields of fundamental and applied research, including mesoscopic physics~\cite{akkermans2007mesoscopic}, biological tissue diagnostics~\cite{wang2007biomedical, tuchin2015tissue}, material characterization~\cite{berne2000dynamic}, atmospheric, oceanic and planetary physics~\cite{stamnes2017radiative} and computer graphics~\cite{pharr2023physically}. The framework of radiative transfer theory~\cite{chandrasekhar1960radiative}, wherein light transport is modeled as an incoherent process, offers a powerful means to investigate the optical properties of scattering and absorbing materials. More precisely, one shows that, when the typical distance between two scattering events is much larger than the wavelength in the medium, ensemble-averaged quantities such as the average luminance, energy density, reflectance, transmittance, absorbance, and so on, can be predicted by simulating the propagation of ``photons'' (to be interpreted as independent, non-interfering light particles) subject to random scattering and absorption events with given statistics.

A standard numerical approach to solve the radiative transfer problem is to perform random-walk Monte Carlo simulations~\cite{carter1975particle, flock1989monte, flock1989monte2, jacques1995monte, novak2018monte} with the scattering and absorption mean free paths as well as the scattering phase function as input parameters. In the simplest case of a dilute, random suspension of particles in a homogeneous background medium, these random-walk parameters can be formally related to the properties of the microstructure (particle size, shape, nature and number density)~\cite{carminati2021principles}. Assuming that the particles are distributed randomly and uniformly throughout the volume of the medium eventually leads to Beer-Lambert's law, which shows that the ballistic component of the average intensity should decay exponentially with the propagation distance over a typical length known as the extinction mean free path (accounting for both scattering and absorption)~\cite{carminati2021principles}. Most light transport Monte Carlo algorithms to date, including those employed to predict the optical properties~\cite{yamada2005spatial, okamoto2013monte, herzog2024monte} and appearance~\cite{lanza2024practical} of cosmetic products, rely on exponentially-distributed step lengths.

In many realistic scenarios, however, the assumption of a uniform distribution of scattering or absorption elements is not reasonable. For instance, particles clustering or multiphase background materials with distinct properties may lead to large spatial fluctuations in the particle density as well as spatial refractive index variations. In such cases, the distribution of step lengths between scattering or absorbing events may deviate quite significantly from a decaying exponential. Research on incoherent transport processes in systems with nonexponential extinction has been developed since the middle of the last century with applications in neutron transport in pebble bed reactors~\cite{behrens1949effect}, light absorption in photosynthesizing cells~\cite{rabinowitch1951photosynthesis, duyens1956flattering}, radiative transfer in planetary atmospheres~\cite{natta1984extinction, varosi1999analytical, davis2004photon} and physically-based rendering~\cite{jarabo2018radiative, bitterli2018radiative}. For a recent review of the literature on nonclassical transport, see Ref.~\cite{deon2022hitchhiker}.

The distinction between so-called quenched and annealed disorder is particularly relevant in the framework of nonclassical transport theory. Quenched (or frozen) disorder describes a heterogeneous medium that does not evolve in time, at least on the typical transport time scale. A major consequence of this is the existence of correlations between successive interaction events (not to be confused with correlations in the spatial arrangement of the scattering elements, though the two are not independent one from another~\cite{vynck2023light}). For instance, a photon experiencing a scattering event just after crossing a large nonscattering region is likely to cross the very same region in the opposite direction soon after. As demonstrated by light transport simulations and experiments in nanoporous ceramics with large macropores~\cite{svensson2014light}, such a behavior can affect significantly the diffusivity of a material. In presence of multiple phases with different refractive indices, successive refraction and reflection events between interfaces of a heterogeneity also lead to a correlation on the photon step lengths and directions. Yet, despite their importance in many situations, such correlations are neglected when considering annealed disorder. To give a simple picture, annealed disorder describes to a heterogeneous medium that would be randomly scrambled after each interaction with a heterogeneity. The medium is thus statistically homogeneous in terms of its scattering properties, while the heterogeneity may alter the step length distribution and angular redistribution of photons within it. Assuming annealed disorder makes an analytical approach to the problem more favorable~\cite{larsen2011generalized, bitterli2018radiative} yet at the cost of a reduced predictive accuracy. Special care also needs to be taken with regard to reciprocity when simulating transport with nonexponential step length distributions in presence of material interfaces~\cite{deon2018reciprocal,binzoni2022monte, tommasi2024anomalous}.

The standard approach for quantitative predictions in heterogeneous materials is to perform light transport Monte Carlo simulations with an explicit description of the microstructure~\cite{boas2002three, ren2010gpu, svensson2013holey, svensson2014light}. This approach was used by us recently to predict the diffuse colors of cosmetic products based on two-phase emulsions~\cite{tran2023physically}. The downside of it is the high computational cost it entails, which is associated to the requirement to repeat simulations over many realizations of the disordered microstructure to obtain statistically meaningful results, as well as each and every time the macroscopic shape or size of the material is varied.

Some strategies have been proposed to alleviate the computational load of simulations in heterogeneous media while maintaining a high accuracy.
In the null-collision model~\cite{elhafi2021three}, nonscattering regions are modeled as producing purely forward-scattering events. This allows making the medium virtually homogeneous with an exponential law between successive scattering (or absorption) events. However, the method cannot handle refractive index discontinuities that lead to reflection and refraction and thus, cannot deal with multiphase materials.
In the field of computer graphics, a multiscale approach for modeling granular materials has been proposed~\cite{meng2015multi}. The method relies on explicit path tracing (i.e., in quenched disorder) at the finest scale, a stochastic ``teleportation'' model accounting for intra-grain transport (numerically computed) and inter-grain propagation (analytically estimated) at an intermediate scale, and the diffusion approximation at the largest scale. The computational load is therefore reduced by a simplified description of transport at the two larger scales. However, step correlations due to specific microstructures are partly neglected at the intermediate scale and the diffusion approximation at the largest scale still relies on exponentially-distributed step lengths.
A similar approach, coined ``quasiannealed'', for treating transport at the intermediate scale had been proposed earlier~\cite{svensson2013holey}, where the random-walk process relied on an estimated step length distribution in the quenched disorder while neglecting step correlations. Predictions of the time-resolved transmission through a medium with fractal heterogeneity were found to deviate quite significantly from quenched simulations, thereby showing the importance of step correlations, at least for dynamic transport quantities, in such materials.

In the present paper, we push the idea of exploiting the knowledge of the step statistics in quenched, heterogeneous media, including effective phase functions and step correlations, to simulate light transport in materials of any shape and size more efficiently. In essence, a database containing random-walk trajectories in an infinite-size heterogeneous medium is built from quenched Monte Carlo simulations in a numerically-generated heterogeneous medium with periodic boundary conditions, and is then used to create new random-walk trajectories in an effectively homogeneous (i.e., annealed) finite-size medium. We apply this method to model light transport in a two-phase material consisting in nonscattering spherical droplets dispersed in a turbid phase with varying scattering properties. Simulations on the reflectance of finite-thickness slabs of heterogeneous materials lead to typical errors on the order of a percent compared to quenched simulations and reductions of the computational time by factors from 2.5 to 7 depending on the microstructure parameters, with no consideration of parallelization. The study of the visual appearance of materials, which is central in various research fields in both academia and industry, may greatly benefit from such a possibility.

The remainder of the paper is structured as follows. In Sec.~\ref{sec:QA-MC-method}, we explain how we build a collection of reference random-walk trajectories using a Monte Carlo method in an explicit (quenched) heterogeneous medium and detail our approach to model light transport in any macroscopic material using this database. In Sec.~\ref{sec:A-QA-Q-simulations}, we compare the predictions of this method over simplistic Monte Carlo simulations in annealed disorder to underline the importance of taking step correlations into account. We show the agreement between the simulated reflectances obtained by our method and by the ground-truth explicit modeling of light transport for heterogeneous monodisperse systems of various nature and for different incident angles. Finally, in Sec.~\ref{sec:discussion-conclusion}, we discuss the potential benefit of our method in terms of computational time as well as possible pathways for future research.

\section{Quasiannealed Monte Carlo method}
\label{sec:QA-MC-method}

\begin{figure}
    \centering
    \includegraphics[scale=0.7]{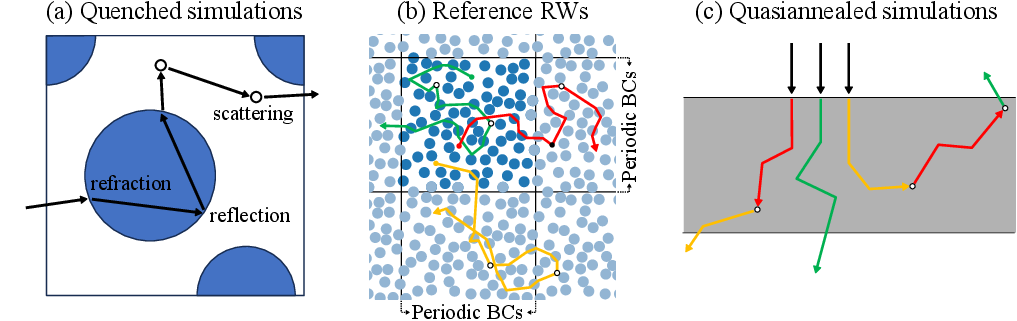}
    \captionsetup{justification=justified, width = \textwidth}
    \caption{From quenched to quasiannealed Monte Carlo simulations for a nonabsorbing heterogeneous medium composed of nonscattering spherical droplets in a turbid medium. (a) 2D illustration of a typical random walk (RW) trajectory in the heterogeneous medium. A random walker may be refracted or reflected when arriving on a droplet surface, and be scattered in the turbid medium between the droplets. The hollow circles indicate scattering events, which occur in the turbid medium only.
    (b) 2D illustration of the simulations in quenched disorder to build up the reference random walk trajectories database. A microstructure is generated by a standard sphere packing algorithm with periodic boundary conditions (BCs). RWs are initialized with random positions and directions in the turbid phase and propagate until a certain number of scattering events is made. (c) 2D illustration of the quasiannealed simulations in a finite-size medium (here, a slab of given thickness). The random-walk trajectories are built from the random-walk trajectory database without an explicit description of the underlying microstructure. A trajectory is terminated when the random walker exits the system either in reflection or in transmission.}
    \label{fig:figMethod}
\end{figure}

In the case of the classical Monte Carlo method for a nonabsorbing homogeneous medium, light propagation is modeled using random walkers whose trajectory is affected by scattering events linked to the properties of the medium: the step length distribution $p(\ell)=1/\ls \exp[-\ell/\ls]$, with the scattering mean free path $\ls$ that represents the average distance between two scattering events, and the scattering phase function $p(\textbf{u} \cdot\textbf{u'})$, which describes the probability to be scattered in a direction $\textbf{u}$ with respect to an incoming direction $\textbf{u'}$. A trajectory is constructed by generating successive step lengths and directions from their probability density distributions until the random walker exits the medium, and the observables (e.g., the reflectance) are obtained by simulating a large number of random walk trajectories. 

The addition of nonscattering regions to the turbid medium affects both the random-walk step lengths and scattering phase function. As illustrated in Fig.~\ref{fig:figMethod}(a) in the case of spherical droplets, light can be refracted or reflected at the droplet boundaries and scattered in the interstitial turbid medium. The former interaction events are modeled here by classical ray tracing in the framework of geometry optics and the latter by a classical Monte Carlo method with exponential step length distribution.

In the following, we first describe how we simulate the propagation of random walkers in the heterogeneous medium while considering the explicit microstructure to build a statistical description of the successive interactions that alter light transport. An infinite-size heterogeneous medium is modeled via the supercell approach, that is, using periodic boundary conditions with a large unit cell [Fig.~\ref{fig:figMethod}(b)]. Then, we detail the method used to model light transport in finite-size materials from this database of trajectories [Fig.~\ref{fig:figMethod}(c)].

\subsection{Reference random walkers in quenched disorder}

We consider a periodic three-dimensional (3D) heterogeneous medium composed of two distinct nonabsorbing phases: (i) a statistically homogeneous turbid phase and (ii) a dispersed phase consisting of a random distribution of nonscattering, nonoverlapping spherical droplets. In the case of the heterogeneous medium previously described, we model light propagation using a Monte Carlo algorithm~\cite{tran2023physically} that explicitly accounts for refraction and reflection events at the surface of the droplets and scattering events in the turbid medium.

We employ the following scheme to account for the heterogeneities in the system: given a tentative step of length $\ell$ and direction $\textbf{u}$ sampled from the properties of the turbid medium, we propagate the random walker until it either intersects a droplet or travels the full required distance. In the latter case, it is similar to the classical Monte Carlo method for which we simulate a new scattering event and reiterate the process. In the former case, we compute the polarization-averaged Fresnel coefficient from the refractive indices of the two phases to account for either the transmission or reflection of the random walker and use Snell-Descartes law to compute the deviated direction~\cite{jackson1999classical}. Note that the random walker may experience multiple reflection events within the droplet before exiting it. Eventually, the step is completed once the random walker has traveled the distance $\ell$ in the turbid phase. Thus, a step in this heterogeneous system is typically longer than $\ls$ and composed of several parts of length $s$ that are correlated to each other.

\begin{figure}
    \centering
    \includegraphics[scale=0.7]{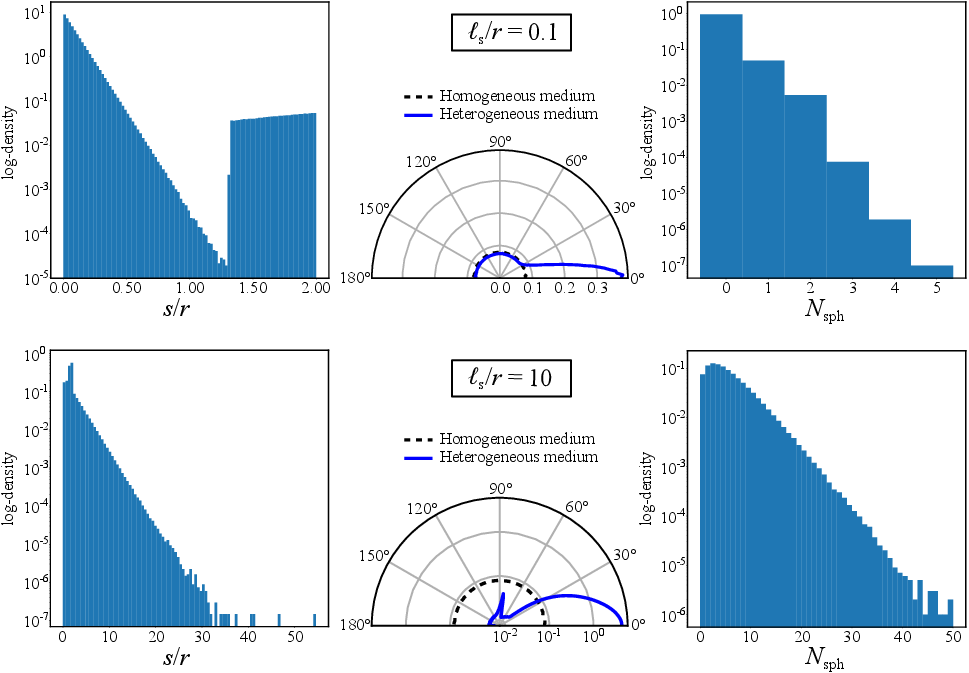}
    \captionsetup{justification=justified, width = \textwidth}
    \caption{Comparison of steps statistics in infinite heterogeneous materials consisting of spherical droplets with radius $r$ and refractive index $n_\text{sph}=1.33$ at a volume fraction $f = 0.3$ in a turbid medium with effective refractive index $n_\text{turb}=1.0$. Two configurations are considered: $\ls/r = 0.1$ (top) and $\ls/r = 10$ (bottom) with fixed scattering anisotropy parameter $g=0$. (\emph{left}) Histogram of the step lengths $s$ between two interaction events (refraction, reflection or scattering) in the heterogeneous material. (\emph{middle}) Phase functions of the statistically homogeneous material (red dashed lines) and of the heterogeneous material (blue solid lines). (\emph{right}) Histogram of the number of spheres crossed between two scattering events. The step statistics depends strongly on the microstructure properties of the material.}
    \label{fig:stepsStatistics}
\end{figure}

Figure~\ref{fig:stepsStatistics} shows the influence of the heterogeneities on the step statistics obtained by propagating random walkers in a large heterogeneous medium with periodic boundary conditions. The spherical droplets have a radius $r$, are placed using a random sequential addition algorithm with a volume filling fraction $f=0.3$, and for the sake of simplicity at this stage, scattering in the turbid phase is assumed to be fully isotropic, $p(\mathbf{u}\cdot\mathbf{u}')=1/(4\pi)$. We compare the situations in which $\ls/r=0.1$ (upper row) and $\ls/r=10$ (lower row). The left panels show histograms of the distance between two interaction events (refraction, reflection or scattering) in the heterogeneous material, the middle panels show the effective phase functions in comparison to that of a homogeneous medium with isotropic scattering, and the right panels show the distribution of the number of spherical droplets crossed between two scattering events. Setting $\ls/r=0.1$ leads to a situation in which a random walker has a relatively low probability of interacting with a droplet between two scattering events. Effectively, this leads to the addition of a low-probability bump (related to the chord-length distribution of a spherical droplet) on top of an exponential decay (related to the turbid phase) in the step length distribution, and to a notable forward-scattering component in the phase function. Setting $\ls/r=10$ leads to the opposite situation, in which a random walker is very likely to interact with one or several droplets between two scattering events. The step length distribution in the droplet is then predominant in the overall step length distribution (which still decays exponentially in the asymptotic limit), and the phase function essentially stems from refraction and reflection events at the droplet interfaces. In this case, transport is driven mostly by the interaction with the disordered arrangement of droplets. All in all, this shows that transport in heterogeneous media can be of very different nature, depending on the microstructure properties.

To address this variety of situations in our approach, we use Monte Carlo simulations in quenched disorder to build a collection of reference random-walk trajectories by randomly generating initial positions and initial directions in the turbid medium and defining an amount of steps to be propagated. For the sake of clarity, we note $N$ the number of reference random-walk trajectories and $M$ the number of scattering events they undergo in the medium. By concatenating all trajectories, we build a database consisting of $N \times M$ independent steps that characterize the properties of the heterogeneous medium: by considering the step as the most basic unit, we ensure that the correlations due to the interaction with one or several droplets between two scattering events are considered fully.

\subsection{Quasiannealed simulations}

In the classical Monte Carlo method for statistically homogeneous media, the random walk process relies on the generation of successive and independent steps from the step length distribution and phase function. Our approach to generate a new step, instead, simply consists in randomly generating, with equal probability, a step from the database. This recorded step contains successive interactions encountered in the medium. However, we cannot directly apply it to propagate a random walker due to the mistmatch between the actual direction $\textbf{D}_\text{in}$ and the recorded direction $\textbf{D}_\text{ref}$. To maintain the global angular distribution properties, we use a transformation called Rodrigues rotation to align these two vectors~\cite{rodrigues1840lois, dai2015euler}. In the general form, given a rotation axis $\textbf{k}$ and angle $\theta$, we compute the transformation of the vector \textbf{v} to the rotated vector $\textbf{v}_\text{rot}$ from 
\begin{equation}
    \textbf{v}_{\text{rot}} = \textbf{v} \cos{\theta} + (\textbf{k} \times \textbf{v}) \sin{\theta} + \textbf{k} (\textbf{k} \cdot \textbf{v}) (1 - \cos{\theta}).
    \label{eq:RodriguesRotation}
\end{equation}
We use the axis-angle pair of parameters $(\textbf{N}, \theta)$ where $\textbf{N}$ is given by the normal to the plane defined by the two vectors $(\textbf{D}_{\text{in}}, \textbf{D}_{\text{ref}})$ and $\theta$ is the angle between these two vectors in this plane. These parameters are computed as 
\begin{eqnarray}
    \label{eq:AxisAngleRotation}
    |\theta| &=& \arccos \left( \dfrac{\textbf{D}_{\text{in}} \: \cdot \: \textbf{D}_{\text{ref}}}{\| \textbf{D}_{\text{in}} \| \| \textbf{D}_{\text{ref}} \|} \right), \\ 
    \textbf{N} &=& \dfrac{\textbf{D}_{\text{in}} \times \textbf{D}_{\text{ref}}}{\left\| \textbf{D}_{\text{in}} \times \textbf{D}_{\text{ref}} \right\|}.
\end{eqnarray}
There exist two solutions for the angle $\theta$, corresponding to clockwise and counter-clockwise rotation around the axis $\textbf{N}$. To solve this ambiguity, we compute both rotations and use the angle that minimizes the reconstruction of vector $\textbf{D}_\text{in}$. The random walk is then obtained by applying the chosen rotation to each component of the recorded step.

In the case of a laterally-infinite slab of finite thickness, the positions of each random walker is monitored after each segment. This allows us to determine whether the propagation process should continue or if the random walker should be considered as exiting the system, either in reflection or transmission. Furthermore, we consider that the dispersed phase, i.e. the spherical droplets, are fully embedded in the turbid phase. Thus, we need to ensure that random walkers exit the medium through the latter. In order to conform to this condition, we monitor the last phase travelled by each random walker and eliminate from the statistics those that exit the medium through the dispersed phase. Then, we propagate random walkers until we reach the predefined total number. This condition emulates a scenario in which the heterogeneities do not intersect the macroscopic object boundary. Note that trajectories going through a droplet that intersects the medium boundary yet without exiting the medium are still considered, which is an approximation we make here.

In summary, Monte Carlo simulations are first performed in a virtually infinite quenched heterogeneous medium. This provides us with a database of steps whose statistics include correlations and angular deviation due to the interaction with the droplets between two scattering events in the turbid phase. We then use the information retrieved from these reference random walkers to simulate light propagation in finite-size macroscopic materials while eliminating the reliance on the explicit knowledge of the microstructure. Consequently, this method can be classified as a \emph{quasiannealed}. It shares similarities with annealed simulations by neglecting the explicit consideration of heterogeneities. However, unlike traditional annealed simulations, correlations are mostly preserved.

\section{Annealed, quasiannealed versus quenched simulations}
\label{sec:A-QA-Q-simulations}

\subsection{Importance of correlations on reflectance}

\begin{figure}
    \centering
    \includegraphics[width = 7cm]{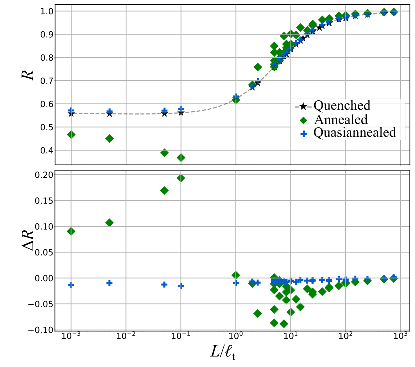}
    \captionsetup{justification=justified, width = \textwidth}
    \caption{Comparison between annealed, quasiannealed and quenched simulations for a monodisperse system of thickness $L = 50 r$ consisting of a volume fraction $f = 0.3$ of spherical droplets with radius $r$ and refractive index $n_{\text{sph}} = 1.33$ incorporated in a turbid medium with refractive index $n_{\text{turb}} = 1.0$. (\emph{top}) Reflection coefficients for varying transport mean free paths. (\emph{bottom}) Reflectance difference with quenched simulations as reference for varying transport mean free paths. The quasiannealed approach provides much better agreement with quenched simulations compared to the annealed simulations.}
    \label{fig:QvsQAvsA}
\end{figure}

It is instructive to start by evaluating the importance of correlations on the reflectance of a finite-thickness heterogeneous material in a practical example. We consider a slab of material with thickness $L$, consisting of a turbid medium with effective refractive index $n_{\text{turb}}$ comprising a random assembly of identical, nonscattering spherical droplets with radius $r$ and refractive index $n_{\text{sph}}$ at a volume filling fraction $f$. The droplets neither overlap each other nor intersect the material interfaces. Here, we set $n_{\text{turb}} = 1.0$, $n_{\text{sph}} = 1.33$, $L/r= 50$ and $f = 0.3$, noting that these parameters will be varied later on, and we assume that the heterogeneous material is placed in an environment of same refractive index as the turbid phase in order to avoid reflection at the interfaces. For convenience, we use the Henyey-Greenstein phase function~\cite{henyey1941diffuse} to model the angular distribution upon a scattering event in the turbid phase. This makes that scattering in the latter is numerically adjusted from the pair of scalar parameters $(\ls, g)$, where $g=\int_{4\pi} p(\mathbf{u} \cdot \mathbf{u}') \mathbf{u} \cdot \mathbf{u}' d\mathbf{u}$ is the scattering anisotropy factor. According to radiative transfer theory, this defines a new length scale $\lt$, known as the transport mean free path, which represents the average distance for a photon to loose memory of its initial propagation direction and is defined as $\lt = \ls/(1-g)$.

The evolution of the total reflectance $R$ predicted from quenched simulations on a finite-thickness slab for normally-incident light and over several orders of magnitude of $L/\lt$ is shown in the top panel of Fig.~\ref{fig:QvsQAvsA}. The reflectance ``ground truth'' is obtained by averaging over 5 independent simulations of {20 000} random walk trajectories for several pairs of scattering parameters $(\ls, g)$. As expected, the reflectance $R$ increases towards 1 with increasing $L/\lt$, corresponding to a material that becomes increasingly more opaque due to multiple scattering in the turbid medium, whereas it saturates to a strictly positive value when $L/\lt$ tends to 0. In this limit, scattering is effectively produced by the disordered arrangement of spherical droplets with a refractive index contrast.

We now attempt to reproduce these predictions with annealed and quasiannealed simulations. The database is built from a total of {20 000} reference random walkers, placed at random initial positions in the turbid phase and propagating with random initial directions for a total of 20 steps, in the corresponding heterogeneous materials with periodic boundary conditions.

As a first attempt, we completely disregard correlations in the propagation distances and deviation angles between each segment of a step. To do this, we simply build a distribution of distances between interaction events (i.e., scattering, refraction and reflection) and a distribution of angular deviations, which are both independent of the position in the material. The heterogeneous medium is therefore assimilated to a homogeneous medium with a nonexponential step length distribution and an effective phase function, as reported in Fig.~\ref{fig:stepsStatistics}. Successive steps are generated as in a classical Monte Carlo simulation in annealed disorder yet using the numerically-generated distributions. The results, obtained by averaging the reflectance $R$ for a series of 5 simulations of {20 000} random walk trajectories for each set of parameters and shown in Fig.~\ref{fig:QvsQAvsA}, reveal significant deviations, especially at low and intermediate values of $L/\lt$, with a maximal absolute error $|\Delta R|$ of about 0.2 for $L/\lt = 10^{-1}$. The reflectance is systematically underestimated for $L/\lt < 10^0$ and overestimated for $L/\lt > 10^0$. This is in sharp contrast with predictions from quasiannealed simulations, in which correlations within a step, that is in-between two scattering events, are preserved. A much better agreement with the predictions from quenched simulations is obtained, with typical errors of the order of a percent on the whole range of parameters ($10^{-3} \leq L/\lt \leq 10^3$).

These first results on the steady-state reflectance of a slab of heterogeneous material emphasize the importance of taking into account correlations between successive interaction events in-between two scattering events in transport simulations. Note that correlations between successive steps, which were previously shown to impact dynamic transport quantities~\cite{svensson2014light}, are neglected here.

\subsection{Prediction accuracy with varying microstructure parameters}

\begin{figure}
    \centering
    \includegraphics[width=7cm]{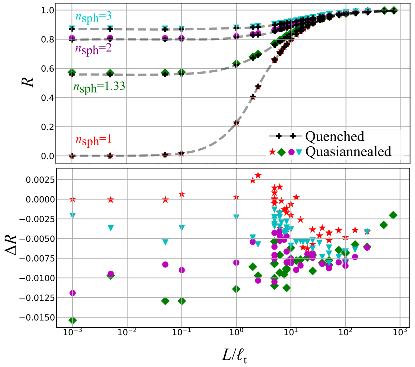}
    \captionsetup{justification=justified, width = \textwidth}
    \caption{Comparative analysis of the reflectance $R$ (\emph{top}) and absolute error $\Delta R$ (\emph{bottom}) between quenched simulations (black crosses) and quasiannealed simulations (colored markers) for light at normal incidence ($\theta_\text{in}=0$) on a heterogeneous material of thickness $L$ containing a disordered ensemble of spherical droplets of radius $r$ at a volume filling fraction $f$ in a turbid medium with scattering parameters $(\ls,g)$. Parameters are $L/r = 50$, $f = 0.3$, $n_\text{turb} = 1.0$. $n_\text{sph}$ is varied from $1$ to $3$ and the results are presented as a function of $L/\lt$. The dashed lines, obtained by a spline interpolation of the values from quenched simulations, serve as guides to the eye. Quasiannealed simulations reproduce well the ground-truth results with a maximum reflectance difference $|\Delta R|$ below 1.6\%.}
    \label{fig:MonoDisperse_n}
\end{figure}

\begin{figure}
    \centering
    \includegraphics[width=7cm]{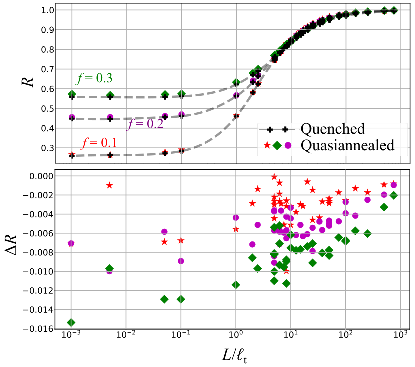}
    \captionsetup{justification=justified, width = \textwidth}
    \caption{Same as in Fig.~\ref{fig:MonoDisperse_n} with $L/r = 50$, $n_\text{sph}=1.33$, $n_\text{turb} = 1.0$ and $f$ is varied from $0.1$ to $0.3$. Quasiannealed simulations reproduce well the ground-truth results with a maximum reflectance difference $|\Delta R|$ below 1.6\%.}
    \label{fig:MonoDisperse_f}
\end{figure}

\begin{figure}
    \centering
    \includegraphics[width=7cm]{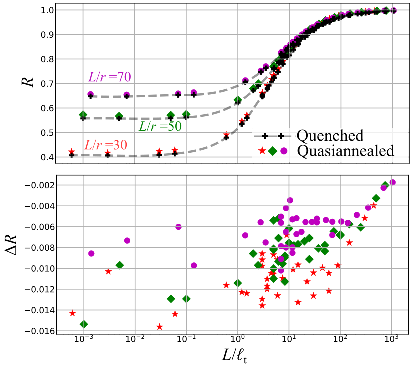}
    \captionsetup{justification=justified, width = \textwidth}
    \caption{Same as in Fig.~\ref{fig:MonoDisperse_n} with $f = 0.3$, $n_\text{sph}=1.33$, $n_\text{turb} = 1.0$ and $L/r$ is varied from $30$ to $70$. Quasiannealed simulations reproduce well the ground-truth results with a maximum reflectance difference $|\Delta R|$ below 1.6\%.}
    \label{fig:MonoDisperse_L}
\end{figure}

We proceed by studying the accuracy of predictions from quasiannealed simulations for different microstructure parameters and light at normal incidence on the material ($\theta_\text{in}=0$). The top panels of Figs.~\ref{fig:MonoDisperse_n}--\ref{fig:MonoDisperse_L} show the computed average reflectances $R$ for 5 iterations of {20 000} random walkers propagating in the system for both quenched and quasiannealed simulations. The signed absolute error $\Delta R$ using quenched simulations as the reference is plotted for the different materials on the bottom panels of Figs.~\ref{fig:MonoDisperse_n}--\ref{fig:MonoDisperse_L}.

First, we vary the refractive index $n_\text{sph}$ of the spherical droplets from 1.0 to 3.0, while keeping the refractive index of the turbid phase fixed to $n_\text{turb}=1.0$. In doing so, we change the respective amount of refracted or reflected light at the interfaces between the two phases. Note that refraction and reflection at the droplet interfaces only depend on the ratio $n_\text{sph}/n_\text{turb}$, so the present results are generally valid for heterogeneous materials with given $n_\text{sph}/n_\text{turb}$, still under the assumption of the present study that the refractive index of the environment is the same as that of the turbid phase. The results are presented in Fig.~\ref{fig:MonoDisperse_n}. When $n_\text{sph} = n_\text{turb} = 1.0$, scattering is only due to the turbid phase, such that $R$ tends to 0 when $L/\lt$ tends to 0. In this regime, light essentially crosses the material without being scattered. The quasiannealed approach naturally captures well this lack of scattering, leading to a vanishingly small error on the reflectance $|\Delta R|$. For higher index contrasts and small values of $L/\lt$, scattering is mostly provided by the droplets, which deviate and reflect more efficiently the incident light, thereby leading to a higher reflectance. Overall, the agreement between quasiannealed and quenched simulations is very good on the considered range of parameters.

Then, we vary the droplet volume fraction $f$ in the turbid medium. This affects the distance between neighboring droplets and consequently the number of droplets that can be crossed between two scattering events. The results are presented in Fig.~\ref{fig:MonoDisperse_f}. Similarly to the previous case, at small values of $L/\lt$, where the light trajectories are mainly affected by the interaction with the droplets, decreasing $f$ leads to a decrease of the reflectance $R$. The material becomes more transparent. As previously, a quantitative agreement is found between the quasiannealed simulations and the quenched simulations on the considered range of parameters.

Finally, we vary the material thickness $L/r$. Because the microstructure parameters remain constant (up to a scaling factor), the quasiannealed simulations here are performed using a single database of random-walk trajectories. The results presented in Fig.~\ref{fig:MonoDisperse_L} show again the predictive accuracy of the quasiannealed approach over a large range of parameters $L/\lt$. They also suggest that quasiannealed simulations are a viable approach to study the optical properties of specific heterogeneous materials with varying macroscopic size and shape.

All in all, we demonstrated the capability of the quasiannealed approach to predict the reflectance of a certain class of heterogeneous materials with varying microstructure properties for normally-incident light with errors on the total reflectance on the order of a percent. This error can be caused either by the fact that correlations between successive steps (i.e., before and after a scattering event in the turbid medium) are ignored, or by our approximate treatment of light transport in the vicinity of the material interfaces. The former is likely negligible in the limit $L/\lt \ll 1$ since a typical trajectory crosses many droplets. The latter may be evidenced by varying the illumination condition, as we shall now see.

\subsection{Prediction accuracy with varying incident angles}

\begin{figure}
    \centering
    \includegraphics[width=7cm]{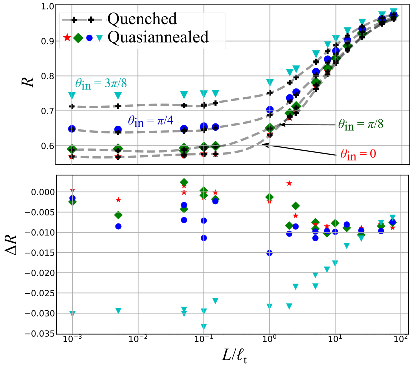}
    \captionsetup{justification=justified, width = \textwidth}
    \caption{Comparative analysis of the reflection coefficient $R$ (\emph{top}) and absolute error $\Delta R$ (\emph{bottom}) between quenched simulations (black crosses) and quasiannealed simulations (colored markers) for light at varying angles of incidence $\theta_\text{in}$ from 0 to $3\pi/8$ on a heterogeneous material of thickness $L$ containing a disordered ensemble of spherical droplets of radius $r$ at a volume filling fraction $f$ in a turbid medium with scattering parameters $(\ls,g)$. Parameters are $L/r = 50$, $f = 0.3$, $n_\text{turb} = 1.0$, $n_\text{sph}=1.33$, and the results are presented as a function of $L/\lt$. The dashed lines, obtained by a spline interpolation of the values from quenched simulations, serve as guides to the eye. A good agreement between predictions from quasiannealed and quenched simulations is observed to incident angles $\theta_\text{in}$ up to $\pi/4$. Deviations are observed at larger angles of incidence with a reflectance difference $|\Delta R|$ of about 3\% for small values of $L/\lt$.}
    \label{fig:MonoDisperse_varAngles}
\end{figure}

To complete the above analysis, we proceed by studying the robustness of the quasiannealed approach by computing the reflectance for various angles of incidence $\theta_\text{in}$ on the heterogeneous material. Results are displayed in Fig.~\ref{fig:MonoDisperse_varAngles} for a heterogeneous material with $L/r = 50$, $f = 0.3$, $n_\text{turb} = 1.0$, $n_\text{sph}=1.33$ as a function of $L/\lt$. Increasing the angle of incidence $\theta_\text{in}$ leads to an increase of the reflectance $R$ in the entire range of parameters $L/\lt$. Similar to transport in statistically homogeneous materials, this can be explained by the first scattering event statistics: at larger incident angles, (i) the path length between the two material interfaces is increased (this length varies as the inverse of the cosine of the angle, $1/\cos \theta_\text{in}$ here), resulting in a higher probability for photons to be scattered at least once; and (ii) the depth at which the first scattering event occurs is reduced (this depth varies with the cosine of the angle, $\cos \theta_\text{in}$ here), resulting in a higher probability to exit the material in reflection by multiple scattering.

Relying on a unique database of random-walk trajectories for the quasiannealed simulations, we find that the accuracy of our predictions tends to decrease at larger incident angles, leading to a reflectance difference $|\Delta R|$ of about 3\% for $\theta_\text{in}=3\pi/8$ (where $R \approx 70\%$). This discrepancy is likely due to our approximate treatment of the material heterogeneity near the interfaces and of how light propagates therein. Recall that our database is constructed from an exploration of an unbounded material, leading to statistically isotropic and translationally-invariant transport properties, whereas the microstructure of finite-thickness slabs is not. In particular, spherical droplets cannot intersect a material interface, such that a photon cannot exit the material through the dispersed phase. We thus expect that, at larger incident angles, a photon should travel on average a longer path in the turbid phase before encountering a heterogeneity than in the unbounded material. Effectively, this leads to an overestimate of the reflectance, which is the behavior observed in Fig.~\ref{fig:MonoDisperse_varAngles}. The same reasoning can be applied to the angular dependence of the reflected light, since light can in principle exit the material at large angles. These results therefore suggest that predictions of the angle-resolved reflectance -- the so-called bidirectional reflectance distribution function (BRDF) -- with the quasiannealed approach might deteriorate with increasing incident and scattered angles. Clearly, light transport near the material interfaces in the framework of quasiannealed simulations is one of the aspects that would deserve further attention in future studies.

\section{Discussion}
\label{sec:discussion-conclusion}

Monte Carlo simulations have been widely used to model light transport in statistically homogeneous media, in which the step length between scattering events is distributed exponentially and successive steps are independent from one another. Strongly heterogeneous media yet lead to nonexponential step length distributions and step correlations. Light transport Monte Carlo simulations in such media therefore rely on an explicit description of the underlying microstructure, which in turn makes the approach very time consuming. In this article, we explored the possibility to exploit the knowledge of typical random-walk trajectories in quenched disorder to simulate light transport in heterogeneous materials without an explicit description of the microstructure. The quasiannealed approach proposed here relies on a two-step process. First, we build a database of random-walk trajectories in a virtually infinite heterogeneous material via quenched Monte Carlo simulations. These trajectories account for nonexponential step length distributions and step correlations. Second, we simulate light transport in finite-size heterogeneous materials by generating new random-walk trajectories from randomly-generated steps in the database. Our hope is that the pre-computation of the database can alleviate the computational cost of quenched simulations, especially when samples of various macroscopic size and shape are investigated.

The accuracy and robustness of our method were tested by comparing predictions from quenched and quasiannealed simulations for the reflectance of slabs of nonabsorbing heterogeneous materials, consisting in a random dispersion of spherical, nonscattering droplets in a turbid medium. Good agreement was found on a large set of structural parameters for normally incident light, with typical absolute errors on the reflectance on the order of a percent. Although discrepancies appeared with increasing angle of incidence, which we explained as a consequence of our treatment of the material heterogeneity and light transport near the material interfaces, the quasiannealed approach appears as a reliable alternative to quenched simulations in terms of accuracy.

The remaining question to address is whether or not a quasiannealed approach can mitigate the computational cost of quenched simulations. First results are presented in Table~\ref{table:ComputationTime}, where we report the times required to compute the reflectance from finite-thickness slabs of heterogeneous materials for different values of $L/\lt$ and at normal incidence, using either quenched or quasiannealed simulations. As previously, the reported values are averages of the reflectance obtained from 5 repetitions of {20 000} random walk trajectories (for both quenched and quasiannealed simulations). The reference database was obtained by propagating {20 000} random walkers in the quenched heterogeneous material for a total of 20 steps between successive scattering events. Both approaches were implemented on Python 3.7 and executed on a virtual machine hosted by the GCP Platform (e2-highmem-16, 16 vCPU, 8 cores) with no consideration of parallelization.

\begin{table}[!ht]
\centering
\begin{tabular}{
 l
 S[table-format=-4.0,
   table-figures-uncertainty=2, 
   table-text-alignment=center,
   table-figures-decimal=0]
 S[table-format=-3.0,
   table-figures-uncertainty=2, 
   table-text-alignment=center,
   table-figures-decimal=0]
 S[table-format=-4.0,
   table-figures-uncertainty=1, 
   table-text-alignment=center,
   table-figures-decimal=0]
 S[table-format=-4.0,
   table-figures-uncertainty=1, 
   table-text-alignment=center,
   table-figures-decimal=0]
}
\toprule
 & \multicolumn{1}{c}{Quenched (s)} & \multicolumn{1}{c}{Quasiannealed (s)} & \multicolumn{1}{c}{Database (s)} \\
\midrule
   $L/\lt = 500$   & 3442 \pm 72 & 1368 \pm 79 & 60 \pm 1\\
  $L/\lt = 50$    & 951 \pm 19 & 275 \pm 13 & 151 \pm 1\\
 $L/\lt = 5$    & 427 \pm 3 & 61 \pm 1 & 1064 \pm 8\\
\bottomrule
\end{tabular}
\captionsetup{justification=justified, width = \textwidth}
\caption{Computational times associated to quenched and quasiannealed simulations for the reflectance of normally-incident light from a slab of heterogeneous material of thickness $L = 50r$ containing a volume fraction $f = 0.3$ of spherical droplets of radius $r$ with refractive index $n_\text{sph} = 1.33$ in a turbid medium with effective refractive index $n_\text{turb}=1.0$. Isotropic scattering ($g=0$) is considered here for the sake of simplicity. The reported values are the average values and the standard deviations obtained from a set of 5 independent simulations of {20 000} random-walk trajectories. (\emph{left column}) Quenched simulations. (\emph{middle column}) Quasiannealed simulations. (\emph{right column}) Database of random-walk trajectories.}
\label{table:ComputationTime}
\end{table}

We observe that the computational time decreases for both quenched and quasiannealed simulations upon decreasing $L/\lt$ (left and middle columns in Table~\ref{table:ComputationTime}), which is because fewer steps on average are required to exit the heterogeneous material, either in reflection or in transmission. In contrast, the time associated to building the database of random-walk trajectories increases (right column in Table~\ref{table:ComputationTime}). The number of spherical droplets crossed by a random walker between two scattering events indeed increases with decreasing $L/\lt$, implying more interactions to compute in each step. This computational time can even be longer than the quenched simulation in the finite-thickness slab, but recall that this initial effort has to be made only once per microstructure and is independent of the macroscopic size and shape of the material. More importantly, quasiannealed simulations lead to a speed-up of about $2.5$ to $7$ times compared to quenched simulations in these examples. Considering further that quasiannealed simulations may be more amenable to parallelization than quenched simulations, these initial tests show the potential benefit of a quasiannealed approach for numerical studies of light transport in heterogeneous materials, especially when materials of different macroscopic size and shape are investigated.

Several improvements and further developments of the method can be conceived at this stage. A first pathway for future studies concerns the optimization of the number and length (in number of steps) of reference random-walk trajectories required to build a database that is statistically representative of light transport in quenched heterogeneous media. A trade-off between accuracy and computational time may be found on the total number of steps to be performed, since random walkers do not explore the heterogeneous material in the same way when the scattering properties of the turbid phase and the volume filling fraction of the dispersed phase vary. Second, we considered here a nonabsorbing heterogeneous material made of monodisperse, nonscattering spherical droplets in a turbid medium, with an effective refractive index equal to that of the ambient medium. It will be necessary in future studies to get closer to real-world systems, including nonspherical, polydisperse heterogeneities, absorbing materials, and real material interfaces with refractive index contrast. Note that absorption can be incorporated in post-processing, provided that the step length distributions in the different phases are recorded in the simulations~\cite{mupparapu2015path}. Importantly, scattering could similarly be incorporated \textit{a posteriori} from quenched Monte Carlo simulations in heterogeneous materials with a transparent phase replacing the turbid phase. This would considerably reduce the initial computational effort associated to the construction of the database of random-walk trajectories to a single computation per type of heterogeneity. Finally, going further along this line, it would be of practical interest to fully replace the numerical construction of the database of random-walk trajectories by a mathematical model accounting for nonexponential step length distributions and step correlations, based on the microstructure properties~\cite{torquato2002random, svensson2013holey, binzoni2023probability, binzoni2024probability}. Eventually, this would enable a fast prediction of the optical properties of arbitrary macroscopic heterogeneous materials and, when associated to an efficient rendering engine, be of great use for appearance prediction and design (e.g., for cosmetic products~\cite{lanza2024practical}).

\begin{backmatter}
\bmsection{Funding}
This study has been funded by L'Or\'{e}al.

\bmsection{Acknowledgments}
The authors acknowledge Mathieu H\'{e}bert and Philippe Lalanne for fruitful discussions at the initial stage of the project, as well as Lorenzo Pattelli for critical feedbacks on our original manuscript.

\bmsection{Disclosures}
LT: L'Or\'{e}al Group (E), BA: L'Or\'{e}al Group (E), KV: L'Or\'{e}al Group (F).

\bmsection{Data Availability Statement}
Data underlying the results presented in this paper are not publicly available at this time but may be obtained from the authors upon reasonable request.


\end{backmatter}


\begin{thebibliography}{10}
\newcommand{\enquote}[1]{``#1''}

\bibitem{akkermans2007mesoscopic}
E.~Akkermans and G.~Montambaux, \emph{Mesoscopic Physics of Electrons and Photons} (Cambridge University Press, 2007).

\bibitem{wang2007biomedical}
L.~V. Wang and H.-i. Wu, \emph{Biomedical Optics: Principles and Imaging} (John Wiley \& Sons, 2007).

\bibitem{tuchin2015tissue}
V.~V. Tuchin, \emph{Tissue Optics : Light Scattering Methods and Instruments for Medical Diagnosis} (SPIE Press, 2015).

\bibitem{berne2000dynamic}
B.~J. Berne and R.~Pecora, \emph{Dynamic Light Scattering: with Applications to Chemistry, Biology, and Physics} (Dover Publications, 2000).

\bibitem{stamnes2017radiative}
K.~Stamnes, G.~E. Thomas, and K.~Stamnes, \emph{Radiative Transfer in the Atmosphere and Ocean, 2nd edition} (Cambridge University Press, 2017).

\bibitem{pharr2023physically}
M.~Pharr, W.~Jakob, and G.~Humphreys, \emph{Physically-Based Rendering: from Theory to Implementation} (MIT Press, 2023).

\bibitem{chandrasekhar1960radiative}
S.~Chandrasekhar, \emph{Radiative Transfer} (Courier Corporation, 1960).

\bibitem{carter1975particle}
L.~L. Carter and E.~D. Cashwell, \enquote{Particle-transport simulation with the {Monte Carlo} method,} Tech. rep., Los Alamos National Lab.(LANL), Los Alamos, NM (United States) (1975).

\bibitem{flock1989monte}
S.~T. Flock, M.~S. Patterson, B.~C. Wilson, and D.~R. Wyman, \enquote{{Monte Carlo} modeling of light propagation in highly scattering tissues. {I.} model predictions and comparison with diffusion theory,} {\protect\JournalTitle{IEEE Trans. Biomed. Engin.}} \textbf{36}, 1162--1168 (1989).

\bibitem{flock1989monte2}
S.~T. Flock, M.~S. Patterson, and B.~C. Wilson, \enquote{{Monte Carlo} modeling of light propagation in highly scattering tissues. {II.} comparison with measurements in phantoms,} {\protect\JournalTitle{IEEE Trans. Biomed. Engin.}} \textbf{36}, 1169--1173 (1989).

\bibitem{jacques1995monte}
S.~L. Jacques and L.~Wang, \enquote{{Monte Carlo} modeling of light transport in tissues,} in \emph{Optical-thermal response of laser-irradiated tissue,}  (Springer, 1995), pp. 73--100.

\bibitem{novak2018monte}
J.~Nov{\'a}k, I.~Georgiev, J.~Hanika, and W.~Jarosz, \enquote{{Monte Carlo} methods for volumetric light transport simulation,} in \emph{Comput. Graph. Forum,}  vol.~37 (Wiley Online Library, 2018), pp. 551--576.

\bibitem{carminati2021principles}
R.~Carminati and J.~C. Schotland, \emph{Principles of Scattering and Transport of Light} (Cambridge University Press, 2021).

\bibitem{yamada2005spatial}
M.~Yamada, M.~D. Butts, and K.~K. Kalla, {\protect\JournalTitle{Int. J. Cosmet. Sci.}} \textbf{27}, 354--354 (2005).

\bibitem{okamoto2013monte}
T.~Okamoto, T.~Kumagawa, M.~Motoda, \emph{et~al.}, \enquote{{Monte Carlo} simulation of light reflection from cosmetic powder particles near the human skin surface,} {\protect\JournalTitle{J. Biomed. Opt.}} \textbf{18}, 061232--061232 (2013).

\bibitem{herzog2024monte}
B.~Herzog, L.~Bressel, S.~Pulbere, and O.~Reich, \enquote{{Monte Carlo} simulations of light transport in sunscreen formulations,} {\protect\JournalTitle{Photochem. Photobiol. Sci.}} \textbf{23}, 1457--1469 (2024).

\bibitem{lanza2024practical}
D.~Lanza, J.~R. Padr{\'o}n-Griffe, A.~Pranovich, \emph{et~al.}, \enquote{Practical appearance model for foundation cosmetics,} in \emph{Comput. Graph. Forum.},  vol.~43 (Wiley Online Library, 2024), p. e15148.

\bibitem{behrens1949effect}
D.~Behrens, \enquote{The effect of holes in a reacting material on the passage of neutrons,} {\protect\JournalTitle{Proc. Phys. Soc. A}} \textbf{62}, 607 (1949).

\bibitem{rabinowitch1951photosynthesis}
E.~I. Rabinowitch, \emph{Photosynthesis and related processes}, vol.~72 (LWW, 1951).

\bibitem{duyens1956flattering}
L.~Duyens, \enquote{The flattering of the absorption spectrum of suspensions, as compared to that of solutions,} {\protect\JournalTitle{Biochim. Biophys. Acta}} \textbf{19}, 1--12 (1956).

\bibitem{natta1984extinction}
A.~Natta and N.~Panagia, \enquote{Extinction in inhomogeneous clouds,} {\protect\JournalTitle{Astrophys. J.}} \textbf{287}, 228--237 (1984).

\bibitem{varosi1999analytical}
F.~V{\'a}rosi and E.~Dwek, \enquote{Analytical approximations for calculating the escape and absorption of radiation in clumpy dusty environments,} {\protect\JournalTitle{Astrophys. J.}} \textbf{523}, 265 (1999).

\bibitem{davis2004photon}
A.~B. Davis and A.~Marshak, \enquote{Photon propagation in heterogeneous optical media with spatial correlations: enhanced mean-free-paths and wider-than-exponential free-path distributions,} {\protect\JournalTitle{J. Quant. Spectrosc. Radiat. Transfer}} \textbf{84}, 3--34 (2004).

\bibitem{jarabo2018radiative}
A.~Jarabo, C.~Aliaga, and D.~Gutierrez, \enquote{A radiative transfer framework for spatially-correlated materials,} {\protect\JournalTitle{ACM Trans. Graph.}} \textbf{37}, 1--13 (2018).

\bibitem{bitterli2018radiative}
B.~Bitterli, S.~Ravichandran, T.~M{\"u}ller, \emph{et~al.}, \enquote{A radiative transfer framework for non-exponential media,} {\protect\JournalTitle{ACM Trans. Graph.}} \textbf{37}, 225 (2018).

\bibitem{deon2022hitchhiker}
E.~{d'Eon}, \emph{A Hitchhiker's Guide to Multiple Scattering} (2022). \url{https://eugenedeon.com/hitchhikers} (accessed 12 Nov. 2024).

\bibitem{vynck2023light}
K.~Vynck, R.~Pierrat, R.~Carminati, \emph{et~al.}, \enquote{Light in correlated disordered media,} {\protect\JournalTitle{Rev. Mod. Phys.}} \textbf{95}, 045003 (2023).

\bibitem{svensson2014light}
T.~Svensson, K.~Vynck, E.~Adolfsson, \emph{et~al.}, \enquote{Light diffusion in quenched disorder: Role of step correlations,} {\protect\JournalTitle{Phys. Rev. E}} \textbf{89}, 022141 (2014).

\bibitem{larsen2011generalized}
E.~W. Larsen and R.~Vasques, \enquote{A generalized linear boltzmann equation for non-classical particle transport,} {\protect\JournalTitle{J. Quant. Spectrosc. Radiat. Transf.}} \textbf{112}, 619--631 (2011).

\bibitem{deon2018reciprocal}
E.~d’Eon, \enquote{A reciprocal formulation of nonexponential radiative transfer. 1: Sketch and motivation,} {\protect\JournalTitle{Journal of Computational and Theoretical Transport}} \textbf{47}, 84--115 (2018).

\bibitem{binzoni2022monte}
T.~Binzoni and F.~Martelli, \enquote{Monte carlo simulations in anomalous radiative transfer: tutorial,} {\protect\JournalTitle{Journal of the Optical Society of America A}} \textbf{39}, 1053--1060 (2022).

\bibitem{tommasi2024anomalous}
F.~Tommasi, L.~Pattelli, S.~Cavalieri, \emph{et~al.}, \enquote{Anomalous radiative transfer in heterogeneous media,} {\protect\JournalTitle{Advanced Theory and Simulations}} \textbf{7}, 2400182 (2024).

\bibitem{boas2002three}
D.~A. Boas, J.~P. Culver, J.~J. Stott, and A.~K. Dunn, \enquote{Three dimensional {Monte Carlo} code for photon migration through complex heterogeneous media including the adult human head,} {\protect\JournalTitle{Opt. Express}} \textbf{10}, 159--170 (2002).

\bibitem{ren2010gpu}
N.~Ren, J.~Liang, X.~Qu, \emph{et~al.}, \enquote{{GPU}-based {Monte Carlo} simulation for light propagation in complex heterogeneous tissues,} {\protect\JournalTitle{Opt. Express}} \textbf{18}, 6811--6823 (2010).

\bibitem{svensson2013holey}
T.~Svensson, K.~Vynck, M.~Grisi, \emph{et~al.}, \enquote{Holey random walks: Optics of heterogeneous turbid composites,} {\protect\JournalTitle{Phys. Rev. E}} \textbf{87}, 022120 (2013).

\bibitem{tran2023physically}
L.~Tran, P.~Lalanne, B.~Askenazi, and K.~Vynck, \enquote{Physically based modeling of the colored appearance of complex media for cosmetics,} in \emph{Proceedings of the 33rd IFSCC Congress,}  (Barcelona, Spain, 2023).

\bibitem{elhafi2021three}
M.~El~Hafi, S.~Blanco, J.~Dauchet, \emph{et~al.}, \enquote{Three viewpoints on null-collision {Monte Carlo} algorithms,} {\protect\JournalTitle{J. Quant. Spectrosc. Radiat. Transf.}} \textbf{260}, 107402 (2021).

\bibitem{meng2015multi}
J.~Meng, M.~Papas, R.~Habel, \emph{et~al.}, \enquote{Multi-scale modeling and rendering of granular materials.} {\protect\JournalTitle{ACM Trans. Graph.}} \textbf{34}, 49--1 (2015).

\bibitem{jackson1999classical}
J.~D. Jackson, \emph{Classical Electrodynamics, 3rd edition} (John Wiley \& Sons, 1999).

\bibitem{rodrigues1840lois}
O.~Rodrigues, \enquote{Des lois g{\'e}om{\'e}triques qui r{\'e}gissent les d{\'e}placements d'un syst{\`e}me solide dans l'espace, et de la variation des coordonn{\'e}es provenant de ces d{\'e}placements consid{\'e}r{\'e}s ind{\'e}pendamment des causes qui peuvent les produire,} {\protect\JournalTitle{J. Math. Pures Appl.}} \textbf{5}, 380--440 (1840).

\bibitem{dai2015euler}
J.~S. Dai, \enquote{{Euler--Rodrigues} formula variations, quaternion conjugation and intrinsic connections,} {\protect\JournalTitle{Mech. Mach. Theory}} \textbf{92}, 144--152 (2015).

\bibitem{henyey1941diffuse}
L.~G. Henyey and J.~L. Greenstein, \enquote{Diffuse radiation in the galaxy,} {\protect\JournalTitle{Astrophys. J.}} \textbf{93}, 70--83 (1941).

\bibitem{mupparapu2015path}
R.~Mupparapu, K.~Vynck, T.~Svensson, \emph{et~al.}, \enquote{Path length enhancement in disordered media for increased absorption,} {\protect\JournalTitle{Opt. Express}} \textbf{23}, A1472--A1484 (2015).

\bibitem{torquato2002random}
S.~Torquato and H.~W. Haslach~Jr, \enquote{Random heterogeneous materials: microstructure and macroscopic properties,} {\protect\JournalTitle{Appl. Mech. Rev.}} \textbf{55}, B62--B63 (2002).

\bibitem{binzoni2023probability}
T.~Binzoni and A.~Mazzolo, \enquote{Probability density function for random photon steps in a binary (isotropic-poisson) statistical mixture,} {\protect\JournalTitle{Scientific Reports}} \textbf{13}, 9887 (2023).

\bibitem{binzoni2024probability}
T.~Binzoni and A.~Mazzolo, \enquote{Probability density functions for photon propagation in a binary (isotropic-poisson) statistical mixture with unmatched positive and negative refractive indexes,} {\protect\JournalTitle{Physical Review E}} \textbf{110}, 054106 (2024).

\end{thebibliography}
\end{document}